\documentstyle[12pt]{article}
\begin{document}

\baselineskip 15pt

\title{\bf About the Notion of Truth in the Decoherent Histories
Approach: \\ a Reply to Griffiths.}
\author{Angelo Bassi\footnote{e-mail: bassi@ts.infn.it}\\ {\small
Department of Theoretical
 Physics of the University of Trieste,}\\ and \\
\\ GianCarlo Ghirardi\footnote{e-mail: ghirardi@ts.infn.it}\\ {\small
Department of
Theoretical Physics of the University of Trieste, and}\\ {\small the Abdus
Salam International
 Centre for Theoretical Physics, Trieste, Italy.}}

\date{}

\maketitle

\begin{abstract}

Griffiths claims that the ``single family rule'', a basic postulate of the
decoherent histories
approach, rules out our requirement that any decoherent history has a
unique truth value,
independently from the decoherent family to which it may belong.
Here we analyze the reasons which make our requirement
indispensable   and we discuss the consequences of rejecting it.

\end{abstract}

\vspace{2cm}

This short letter is a reply to Griffiths' article {\it Consistent
histories, quantum truth functionals, and hidden variables}
\cite{gri1}, in which he has raised some objections to our paper {\it Can
the decoherent
histories description of reality be considered satisfactory?} \cite{bg1}. For
a more detailed analysis of the arguments of \cite{bg1}, we refer the
reader to \cite{bg2}.

First of all, we would like to summarize the main features of the
DH approach, about which there seems not to be a disagreement
between Griffiths and us:

\begin{enumerate}
    \item Within a given decoherent family everything goes like
    in Classical Mechanics: the probability distribution assigned
    to the histories obeys the classical probability rules; it is
    possible to define a Boolean structure, so it is possible to
    speak about the conjunction, disjunction of two histories and about
    the negation of an history; moreover, one can define
    the logical implication between two histories, so that also
    reasonings of the type ``if ... then ...'' are possible.

    \item As Griffiths admits in the above
    quoted paper, it is possible to assign truth--values to
    all histories of a given decoherent family. This move has an
    important physical meaning: it means that, in spite of the probabilistic
structure of the theory,
    one can speak of the properties
    actually possessed by the physical system under study, and not
    only of the probability that such properties be possessed. In order to
understand
    this important point, let us remember that also in
    Classical Statistical Mechanics one generally has only a
    probabilistic knowledge of the physical system; despite of
    this, he can claim that the system has well defined physical
    properties (positions and momenta of its constituents, from which all other
    properties can be derived), but he doesn't know which they are
    simply because he is ignorant about the precise state of the
    system. From the logical--mathematical point of view, the legitimacy of
considering
properties as objectively possessed is
     a consequence of the fact that one can define
    a Boolean algebra in phase space and attach
    truth--values to its subsets in a consistent way. \\
    In Standard Quantum Mechanics, on the other hand, one cannot
    even think that systems possess physical properties prior to
    measurements: mathematically, this is reflected in the
    peculiar properties of the Hilbert space (with dimension greater
    than 2): the set of projection operators
    cannot be endowed with a Boolean structure, and it is not
    possible to attach consistently truth--values to them,
    as implied by the theorems of Gleason, Bell and Kochen and Specker. \\
    Thus, {\it giving a truth value to the histories of a given
    decoherent family corresponds to the assertion that such histories
speak of specific
    physical properties that the system under study
    possesses objectively, independently from our (in general) probabilistic
    knowledge of the system and of any act of measurement}. This, in our
opinion, is the
nicest feature  of the DH formalism, the one emboding all its advantages
with respect to the
standard quantum formalism.
\\

    \item When one deals with more that one decoherent family,
    things become rather problematic: if such families can be
    accomodated into a single decoherent family, then all what we have
    said previously remains valid. If this is not possible (and
    this is likely to happen most of the times), then any
    reasoning, any conclusion derived by using  histories
    which  belong
    to incompatible families, are devoid of any physical meaning.
    Griffiths felt the necessity to promote this fact , which we have
indicated as the
    ``single family rule'', to a basic
    rule of the DH approach: {\it a meaningful description of a
    (closed) quantum mechanical system, including its time
    development, must employ a single framework} [i.e. decoherent
    family] \cite{gri2}. This rule gives rise to some curious situations,
    which do not have any classical analogue, but we will not discuss
    these matters now. Actually, we agree that they do not lead to formal
inconsistencies.
\end{enumerate}

Now, let us come to our argument. The formalism of DH implies, as it is
obvious and can
be easily checked,
 that any given
decoherent history belongs in general to many different decoherent
families. As we have argued under 2., in any of these decoherent
families, such a history has a precise truth--value. As already mentioned, also
Griffiths seems to agree on this. Now the relevant question is: {\it does
the truth value of the considered history depend on the decoherent family
to which it may belong?} We think that the answer must be ``no'',
because (as we said in 2.) truth--values refer to properties {\it
objectively} possessed by the physical system under study, and if
the truth--value of a decoherent history would change according to the
decoherent
family to which it belongs,  also the properties that
such a history attaches to the physical system would change by changing the
decoherent family.
We have formalized these considerations in the following assumption
(which is  assumption (c) of \cite{bg1}):
\begin{quote}
        {\small Any given decoherent history has a unique truth value (0 or
        1), which is independent from the decoherent
        family to which the history is considered to belong.}
\end{quote}

As mentioned in the abstract, in \cite{gri1}, Griffiths claims that such an
assumption violates
the ``single family rule'' and as such it cannot be considered as
part of the DH approach. With reference to this point  we want
first of all to make clear that nowhere, in the original formulations of
the ``single family rule'', it was mentioned that a given
decoherent history can (or cannot) have different truth values
according to the family to which it belongs; nowhere, directly or
indirectly,  reference was made  to our assumption (since we have been the
first to put it
forward). Thus, it is not correct to claim that such a rule already
excluded our assumption. If
Griffiths claims that the ``single family rule'' excludes (c), then he is
proposing a new,
extended, interpretation of such a rule.

Having clarified this point,  we are ready to accept that Griffiths rejects our
assumption: he is perfectly free to do so. But  we  pretend that  he
accepts all the
consequence (which we are going to analyze) of such a move. Denying (c)
simply means to assert
that:
 \begin{quote}
    {\small There are decoherent histories whose truth--values
    depend on the decoherent family to which they (are thought or
considered to)
belong, i.e. in
    some families they are, for example, true, while in other
    families they are false.}
 \end{quote}
 This, in turn, means accepting that statements like ``this table is here'',
 ``the Earth is moving around the Sun'', ``that electron has spin up
 along such a direction'' are --- in general ---
 neither true nor false {\it per se}: each of them
 acquires a truth value only when it is  considered a member of a precise
(among the infinitely
many ones which are possible)
 decoherent family; moreover, their truth values may change according to
 the decoherent family to which they are associated. In some families
 it may be true that ``this table is here'' or that ``the Earth is
 moving around the Sun'', while in other families it may be false that
 ``this table is here'' or that ``the Earth is moving around the Sun''.
 This state of affairs is the direct consequence of denying our
 assumption, and it should be evident to anyone that if one takes such a
position then he is
spoiling the statements of the DH approach of any physical meaning whatsoever.

We can then summarize the whole debate between Griffiths and us in the
following terms. In  our papers [2] and [3], we have considered the
following four assumptions:

\begin{enumerate}
\item[(a)] Every {\it family} of decoherent histories can be (naturally)
endowed with
a Boolean structure allowing to recover classical reasoning,

\item[(b)] Within every decoherent {\it family} it is possible to assign
to its
histories truth values which
preserve the Boolean structure (i.e. they form an homomorphism),

\item[(c)] Every decoherent {\it history} has a unique truth value,
independently from the
decoherent family to which it may be considered to belong,

\item[(d)] Any decoherent {\it family} can be taken into account,
\end{enumerate}
and we have shown that they lead to a Kochen-and-Specker-like
contradiction. This implies that at least one of them must be rejected in
order to avoid
inconsistencies within the DH approach. Griffiths rejects assumption (c),
while we, in
accordance with the previous analysis, believe that this move is
unacceptable. Accordingly, in
our papers we have  suggested that one should limit, resorting to precise
and physically
meaningful criteria, the set of decoherent histories which can be taken into
account. Such a move might lead to a physically acceptable and sensible new
formulation of the
DH approach.

\end{document}